%% file: usenix.tex
\documentclass[letterpaper,twocolumn,10pt]{article}

\usepackage{usenix2019_v3.1}

\usepackage{latexsym,fancyhdr,url}
\usepackage{enumerate}

\usepackage[ruled,vlined,linesnumbered]{algorithm2e}
\usepackage{algorithm2e}
\usepackage{algpseudocode}
\usepackage{booktabs}
\usepackage{xparse} 
\usepackage{xspace}
\usepackage{multirow}
\usepackage{balance}
\usepackage{makecell}
\usepackage{pifont}
\usepackage{enumitem}
\usepackage{xcolor}
\usepackage{colortbl}
\algnewcommand{\algorithmicand}{\textbf{ and }}
\algnewcommand{\algorithmicor}{\textbf{ or }}
\algnewcommand{\OR}{\algorithmicor}
\algnewcommand{\AND}{\algorithmicand}
\algnewcommand{\var}{\texttt}
\definecolor{light-gray}{gray}{0.95}

\usepackage{xcolor}
\usepackage{caption}
\definecolor{codegreen}{rgb}{0,0.6,0}
\definecolor{codegray}{rgb}{0.5,0.5,0.5}
\definecolor{codepurple}{rgb}{0.58,0,0.82}
\definecolor{backcolour}{rgb}{0.95,0.95,0.92}
\usepackage{listings}
\usepackage{caption}

\usepackage{xcolor}
\definecolor{codepurple}{rgb}{0.5,0.0,0.5}

\definecolor{codegreen}{rgb}{0,0.6,0}
\definecolor{codegray}{rgb}{0.5,0.5,0.5}
\definecolor{codeblue}{rgb}{0,0,1}
\definecolor{backcolour}{rgb}{0.99,0.99,0.99}

\lstdefinestyle{cpp}{
    backgroundcolor=\color{backcolour},   
    commentstyle=\color{codegray},
    keywordstyle=\color{codeblue},
    numberstyle=\tiny\color{codegray},
    stringstyle=\color{codepurple},
    basicstyle=\ttfamily\footnotesize,
    breakatwhitespace=false,         
    breaklines=true,                 
    captionpos=b, 
    numbers=left,
    numbersep=5pt,
}
\makeatletter
\renewcommand\paragraph{\@startsection{paragraph}{4}{\z@}%
  {0.1\baselineskip }%
  {-0.5em}
  {\normalfont\normalsize\bfseries}}
\makeatother

\newcommand{\code}[1]{\lstinline[language=c]{#1}}
\newcommand{\sys}{\textsc{DirtyPage}\xspace}

\usepackage{tikz}
 
\newcommand*\halfcirc[1][1ex]{%
  \begin{tikzpicture}
  \draw[fill] (0,0)-- (90:#1) arc (90:270:#1) -- cycle ;
  \draw[thick] (0,0) circle (#1);
  \end{tikzpicture}}

\newcommand*\fullcirc[1][1ex]{\tikz\fill (0,0) circle (#1);} 

\makeatletter         
\def\@maketitle{
\newpage
\null
\vskip -20em 
\begin{center}
\let \footnote \thanks 
{\LARGE \@title \par}     
\vskip -2em
{\large
 \lineskip .5em 
\begin{tabular}[t]{c} 
\@author
\end{tabular}\par} 
\vskip -10em 
{\large \@date} 
\end{center}
\par
\vskip 1.5em}
\makeatother

\usepackage{titlesec}

\titlespacing*{\subsection}
{0pt}{2ex plus 1ex minus .2ex}{1ex plus .2ex}

\usepackage{titlesec}
\titlespacing*{\subsection}
{0pt}{2ex plus 1ex minus .2ex}{1ex plus .2ex}
\titlespacing*{\section}
{0pt}{1.5ex plus 1ex minus .3ex}{1ex plus .1ex}

\date{}

\title{\Large \bf  Take a Step Further: Understanding Page Spray in Linux Kernel Exploitation\vspace{50pt}}
\author{
{\rm Ziyi Guo$^{*}$, Dang K Le$^{*}$, Zhenpeng Lin$^{*}$, Kyle Zeng$^\dagger$} \\
{\rm Ruoyu Wang$^\dagger$, Tiffany Bao$^\dagger$, Yan Shoshitaishvili$^\dagger$, Adam Doupé$^\dagger$, Xinyu Xing$^{*}$} \\
$^{*}$ Northwestern University \\
\textit{\{n7l8m4,dangle2029,zplin\}@u.northwestern.edu} \\
\textit{xinyu.xing@northwestern.edu} \\
$\dagger$ Arizona State University \\
\textit{\{zengyhkyle,fishw,tbao,yans,doupe\}@asu.edu} \\
}

\begin{document}
\maketitle

\pagestyle{empty}

\subsection*{Abstract}
\input{abstract}

\input{sections/1-intro}
\input{sections/2-background}
\input{sections/3-threat_model}

\input{sections/4-exploit_model}
\input{sections/5-callsite_model}

\input{sections/6-static_analysis_model}
\input{sections/7-evaluation}

\input{sections/8-refurbish}
\input{sections/9-defense}

\input{sections/10-related_work}
\input{sections/11-conclusion}
\input{sections/ack}

\bibliographystyle{plain}
\bibliography{ref}

\appendix
\input{sections/12-appendix}
\end{document}

%% file: abstract.tex
Recently, a novel method known as Page Spray emerges, focusing on page-level exploitation for kernel vulnerabilities. Despite the advantages it offers in terms of exploitability, stability, and compatibility, comprehensive research on Page Spray remains scarce. Questions regarding its root causes, exploitation model, comparative benefits over other exploitation techniques, and possible mitigation strategies have largely remained unanswered. In this paper, we conduct a systematic investigation into Page Spray, providing an in-depth understanding of this exploitation technique. We introduce a comprehensive exploit model termed the \sys model, elucidating its fundamental principles. Additionally, we conduct a thorough analysis of the root causes underlying Page Spray occurrences within the Linux Kernel. We design an analyzer based on the Page Spray analysis model to identify Page Spray callsites. Subsequently, we evaluate the stability, exploitability, and compatibility of Page Spray through meticulously designed experiments. Finally, we propose mitigation principles for addressing Page Spray and introduce our own lightweight mitigation approach. This research aims to assist security researchers and developers in gaining insights into Page Spray, ultimately enhancing our collective understanding of this emerging exploitation technique and making improvements to the community.

%% file: sections/1-intro.tex
\section{Introduction}

The Linux operating system, known for its complexity, has gained widespread popularity in today's world. It serves as the foundation for critical real-world components, including cloud systems, infrastructure web servers, and various customized operating systems. 

To defend Linux against potential threats, developers and designers have implemented a range of protective and mitigation techniques, such as SMEP~\cite{smep}, SMAP~\cite{smap}, and KASLR~\cite{kaslr}, operating at different levels. In response to these defensive measures, security researchers have innovatively introduced new exploitation techniques and made assessments, such as Elastic Objects~\cite{elastic}, K(H)eaps~\cite{kheaps}, and DirtyCred~\cite{dirtycred}, to circumvent these security "barriers" during exploitation. 

As the field of exploitation techniques continues to evolve, there remains an ongoing discourse on refining and enhancing control over kernel-heap granularity. Among these techniques, Linux Kernel Heap Object Spray stands out as the predominant method. Heap Object Spray has found utility in exploiting a variety of common vulnerabilities, including Use-After-Free (UAF)~\cite{UAF}, Double Free (DF)~\cite{DF}, and Out-of-Bounds (OOB)~\cite{OOB-write,OOB-read} vulnerabilities. These traditional approaches primarily center around the manipulation of heap object granularity.

In recent times, a page-level exploitation technique has made its appearance in high-quality exploits, which differs from traditional object spray in control granularity. The capability to control and execute data spraying at the page level introduces new possibilities, not only for offensive security researchers but also for defense and mitigation designers approaching security challenges from various angles.

Despite being an emerging technique in Linux kernel exploitation, the security impact of Page Spray has been underestimated, and Page Spray has been considered obsolete to modern Linux systems.
Prior Page Spray attacks~\cite{wenxu}\footnote{This paper names Page Spray physmap-based attacks. We use the term ``Page Spray'' in this paper since this name is more broadly used in the Linux kernel security community.} only work under constrained memory environments (hundreds of MBs memory) where the Linux operating system will allocate kernel-freed physical pages for user-space memory requests when the system is under memory pressure.
Such situation unlikely happens in modern Linux systems as memory increases significantly and that modern Linux operating systems will kill a memory-intensive user-space process to handle extra user-space memory requests under memory pressure.
Therefore, prior Page Spray attacks no longer work in modern Linux operating systems.

However, we discover that Page Spray can still work through new methods. The original page attack relies on user pages for spraying, which will not work in currently normal scenarios due to the separate allocation of user pages and SLUB pages in distinct memory zones. Previous research indicated that during periods of intense memory pressure, the page allocator might allocate user pages to the kernel page zone, facilitating page-level reuse. However, this practice is no longer feasible due to the increase in device memory, posing a risk of destabilizing the kernel and triggering OOM kill. In contrast, the method in our study works effectively under current standard conditions without stressing kernel memory. Instead of leveraging user-space memory allocation, it uses kernel-space callsites to invoke Page Spray within the kernel. 

Furthermore, the contribution of our work is \textbf{not only to analyze the methodologies of Page Spray, but also is a comprehensive study on many aspects of it that have not been studied before}, such as its root cause, exploitation effectiveness \& stability, and how it can be mitigated with existing approaches as well as our own proposed approach.

To this end, we systematically unravel and formalize its exploitation model, accounting for various scenarios and circumstances.

Moreover, we delve into the origins of this technique by scrutinizing kernel code across different subsystems, ultimately identifying three root causes: Raw Page-Level Buffer, Non-linear Page Frags Buffer, and Mmap \& Zero Copy Calls. These root causes unveil inherent security risks at the design level within the Linux Kernel. We propose an analysis model and develop an LLVM-based analyzer to facilitate the identification of potential and potent Page Spray invocations within the Kernel. Leveraging these analysis tools, we successfully pinpoint 21 callsites within the Kernel that can be used in Page Spray exploitation. 

Additionally, we demonstrate Page Spray's distinctive advantages, highlighting its unique attributes concerning exploitability, stability, and compatibility through real-world evaluations. Furthermore, we delve into the principles of mitigating Page Spray, introducing our lightweight mitigation approach. Our research uncovers intriguing insights, such as Page Spray's ability to transform seemingly challenging-to-exploit or even previously unexploitable vulnerabilities into viable targets for security researchers. At a practical level, \textbf{we have successfully applied this technique to various contexts, including Desktop Ubuntu, Android Kernel~\cite{badiouring}, and Cloud Environment in Google kCTF~\cite{kCTF}, won the novel exploitation award, and some of which have involved zero-day exploits.}

In summary, this paper makes the following contributions:

\begin{itemize}[leftmargin=*, itemsep=1pt, topsep=1pt, partopsep=1pt, parsep=1pt]

\item We provide a well-structured and comprehensive formalization of the Page Spray exploitation technique. Our model showcases the compatibility and adaptability of Page Spray across various vulnerability types.

\item We conduct a methodical evaluation of the exploitability and stability of the Page Spray exploitation technique. This emerging and promising technique has been relatively unexplored, making our study a valuable addition to the understanding of Page Spray.

\item We systematically analyze the root causes of Page Spray Exploitation within the Linux Kernel. Our analysis identifies key design-level security risks, categorizing these into three distinct root causes. We also introduce an analysis model, and implement it as a LLVM-based analyzer for Page Spray to identify the occurrences of such root causes in the kernel.

\item To raise awareness within the community, we discuss and evaluate general and potential mitigation techniques for countering Page Spray exploitation. Additionally, we introduce a lightweight mitigation approach.

\end{itemize}

%% file: sections/2-background.tex
\section{Background}
\label{sec:bg}

\subsection{Kernel Heap Allocator}
To enhance memory performance and reduce fragmentation, the Linux kernel has incorporated the concept of an object-based allocator for managing kernel heap objects. 

Among various heap allocators\cite{SLAB,SLOB,SLUB}, the SLUB allocator is the most common one in Linux. From a higher-level perspective, the heap allocator in the Linux kernel can be conceptualized as a cache size-based allocator, closely linked to the page allocator, which will be comprehensively discussed in the subsequent subsection. The ``cache'' serves as the meta-structure for managing various heap objects within the allocator. At the cache granularity level, the kernel allocates memory pages from the page allocator and divides them into uniform-sized fragments, each serving as a memory slot for maintaining a heap object. Throughout the cache's lifecycle, as the available memory within a cache is being exhausted, the kernel allocates additional pages from the page allocator and assigns them to the heap allocator to supplement the memory pool. Conversely, when a specific cache is no longer in use, i.e. there are no more in-use objects within the cache, the kernel prioritizes the recycling of allocated pages, returning them to the page allocator for potential reuse.

\subsection{Kernel Page Allocator}
Kernel page allocator serves as the underlying foundation for kernel heap allocator. This interaction comes into play when the heap allocator either recycles or reclaims pages. The buddy system, which has emerged as a prominent algorithm for memory allocation within the page context, operates by dividing the available memory into blocks. These blocks adhere to a uniform size, with each block possessing dimensions that are exact powers of two. 

Linux kernel implements distinct memory zones, each carrying specific attributes aimed at optimizing page allocation. Within each zone, a reserve of free memory is maintained to facilitate page allocation. In free memory areas, free pages are organized into separate linked lists based on their size and attributes, with each list representing pages of a specific size and attribute. When a page allocation request is initiated, the kernel identifies the most suitable match of free pages by considering the allocation attributes and the required size.

Moreover, the node within the \code{kmem_cache} serves as a structural component monitoring partially full and full slabs within a specific node in a NUMA system. For instance, if a newly allocated object is the final available object within the CPU slab, then the associated ``active'' slab is transferred to the full list corresponding to its node. Simultaneously, the first slab from the CPU slab's partial list is newly designated as the "active" slab.

\paragraph{GFP (Get Free Pages) flags.} \label{sec:gfpflags} GFP flags serve as a means to characterize the attributes of page allocation. Different GFP flags, assigned by higher-level components, have the potential to yield different fundamental behaviors. For example, when utilizing \code{GFP_DMA} and \code{GFP_DMA32} flags, the kernel is instructed to ensure that the allocated memory is accessible by hardware with limited addressing capabilities in DMA.

\paragraph{Fallback Operation.} During the allocation process, beginning at the zone level, the kernel diligently scrutinizes the selected memory zone to ensure a close match with the desired attributes. A comprehensive evaluation is conducted to determine the suitability of the chosen zone for the allocation. In instances where the assessed zone is found to be unsuitable, the memory allocator may proceed to revert or fallback to alternative memory zones for allocation. 

%% file: sections/3-threat_model.tex
\section{Threat Model}

\begin{figure*}[t]
\centering
\includegraphics[width=1\textwidth]{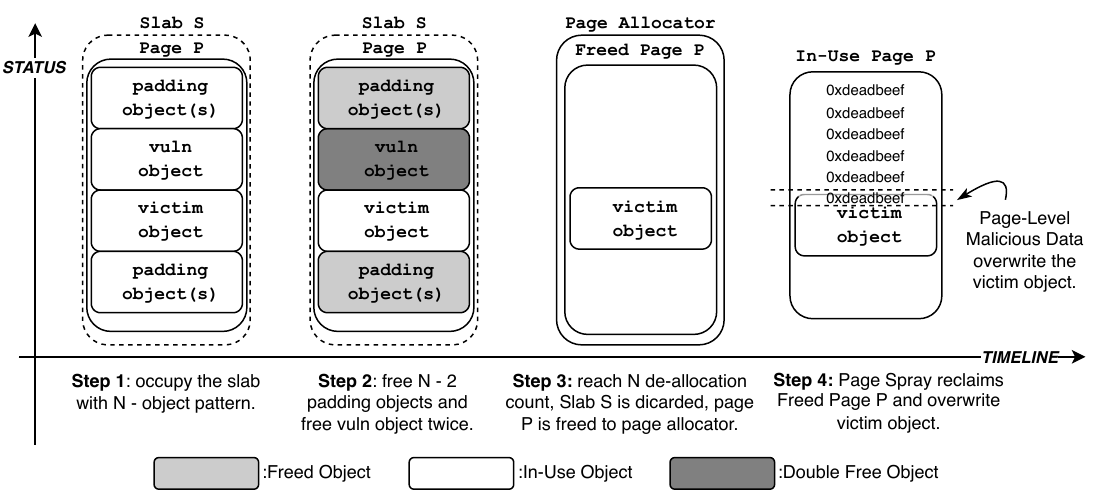}
\vspace{-2px}
\caption{Page Spray Exploit Model for Double Free.}
\label{fig:exploitmodel}
\vspace{-2px}
\end{figure*}

In our established threat model, we assume that, in a real-world exploitation scenario, an adversary gains access to unprivileged permissions within the Linux environment and aims to exploit a kernel vulnerability related to heap memory corruption such as UAF, DF, and OOB as a means to escalate their privileges. It is important to note that we take into consideration typical security protections and mitigations that are enabled on most real-world Linux systems, including KASLR~\cite{kaslr}, SMEP~\cite{smep}, SMAP~\cite{smap}, KPTI~\cite{kpti}, FREELIST RANDOM~\cite{freelistrandom}, HARDENED FREELIST~\cite{freelistharden}, CFI~\cite{cfi}. For Android cases, we also include KCFI~\cite{kcfi}, and Memory Tagging~\cite{mte} on Pixel devices.

Mitigations that are not integrated in the mainline Linux kernel at the time this paper was written, such as AUTOSLAB and SLAB VIRTUAL, are not included in our threat model. However, to provide a comprehensive study, their effectiveness against Page Spray's exploitability is still studied, dicussed and analyzed in Section \S\ref{sec:securepagereuse}.

Furthermore, we make the explicit choice to exclude any hardware-level vulnerabilities or side-channel mechanisms that might be leveraged to facilitate the exploitation process. Finally, we do not account for variations in architecture codes within the Linux kernel as part of our analysis.

%% file: sections/4-exploit_model.tex
\section{Page Spray Exploit Model Analysis}

\label{sec:expmodel}
In this section, we provide a high-level overview of the exploitation model associated with Page Spray, drawing from our extensive analysis. We break down the Page Spray technique into four main steps, showing how it operates during the exploitation of kernel heap vulnerabilities. Furthermore, we showcase the versatility of Page Spray by demonstrating its applicability to different types of vulnerabilities.

\subsection{Basic Exploitation Model}
We now introduce a foundational exploitation model for Page Spray, which we denote as \sys. This model is presented at a high level to offer a clear understanding without delving into excessive technical intricacies. 

\sys operates by employing three key types of heap objects: (1) \textit{padding objects} are responsible for occupying specific positions in the heap memory, (2) \textit{vulnerable object} corresponds to the case-specific object capable of triggering the vulnerability, (3) \textit{victim object} represents the object that is intended to be corrupted by the end of \sys. In our model, \textit{padding objects} and \textit{victim object} can be chosen independently of the vulnerability. However, in order to easily adapt \sys to different vulnerabilities and increase stability, we choose \textit{padding objects} that have controllable size and can be sprayed rapidly on a desired slab cache (such as \code{msg_msg}, \code{iovec}, \code{user_key_payload}, etc.). 

Much like a UAF vulnerability that results in a dangling reference to a freed heap object, \sys aims to achieve a dangling reference to a freed memory page from a reference to the \textit{victim object}, leveraging the primitives provided by the \textit{vulnerable object}, by triggering vulnerabilities such as Double Free, Invalid Free, UAF, and others. In Figure~\ref{fig:exploitmodel}, we illustrate the \sys model for exploiting a Double Free vulnerability. 

In the first step, we initiate a series of allocations to occupy a slab $S$ in a designated slab cache with a specific pattern of objects: \textit{padding objects} - \textit{vulnerable object} - \textit{victim object} - \textit{padding objects}, in a way that the \textit{vulnerable object} and the \textit{victim object} reside in the same page $P$. To achieve this arrangement, we spray \textit{padding objects} twice - once before the allocation of the \textit{vulnerable object} and \textit{victim object}, and once after. Suppose there are $N$ objects per slab in this cache, we now have $S$ populated with one \textit{vulnerable object}, one \textit{victim object}, and $N - 2$ \textit{padding objects}.

To achieve a dangling reference to a freed page $P$, we must ensure that $S$ is to be discarded so that all the memory pages it contains are returned to the page allocator. This requires the deallocation of exactly $N$ objects within $S$, while still maintaining a reference to the \textit{victim object}. We accomplish this by initially freeing all $N - 2$ \textit{padding objects} and then exploiting the Double Free vulnerability to free the \textit{vulnerable object} twice. This operation results in the slab allocator recognizing two separate deallocations, coupled with the $N - 2$ deallocations of the \textit{padding objects}, resulting in a total of $N$ deallocations within $S$. Importantly, the \textit{victim object} remains unaffected and its reference is preserved.

Subsequently, $S$ is discarded, initiating the page recycling process, which returns $P$ within $S$ to the page allocator. Following the recycling of pages, we initiate page-spraying operations from the user space that are capable of triggering direct page allocations (they will be referred to and explained in Section \S\ref{sec:rootcause} as \textit{page-spraying callsites}). Simultaneously, in the kernel space, corresponding page allocations are substantially sprayed, effectively reclaiming the slab pages freed in the previous step, including $P$. This operation facilitates the direct injection of malicious page-level data into the kernel. Notably, regarding $P$, the maliciously crafted page-level data overwrites the data within the \textit{victim object}. For instance, if the \textit{victim object} is a \code{pipe_buffer}, the adversary gains control over the \code{ops} field, which is a pointer to its operations function table.

\subsection{Model Variants}


In different scenarios, the \sys technique can adapt to various variants. When dealing with a UAF vulnerability, it is possible to optimize the model by having the \textit{victim object} and the \textit{vulnerable object} as a single object. This optimization arises from the fact that the UAF object can be inherently freed, eliminating the need for deceiving the allocator and simulating the freeing of all objects. Instead, we can efficiently free all the objects within the slab and still naturally maintain a reference to the freed UAF object.

Moreover, straightforward privilege escalation can be achieved (we achieved this in CVE-2022-2588~\cite{CVE-2022-2588}). To accomplish this, we make adjustments to the specific data used in the Page Spray process, focusing on the \code{cred} object. After constructing a fake \code{cred} object in user space, during the page reclamation step, we employ Page Spray to inject a multitude of these fake \code{cred} objects, overwriting the real credentials in memory pages by writing to pipe pages. Once the target credential is hijacked, the privilege escalation process succeeds.

Finally, cross-cache technique can also be integrated into the \sys model as a plug-in component, depending on the specific object being exploited or the nature of the \textit{vulnerable object}.

\subsection{Model Adaptability}
The model described in the preceding subsections is noteworthy because it relies solely on the fundamental nature of the vulnerability itself. Whether it involves an extra free operation in the case of a Double Free or Invalid Free bug, or a dangling reference to a freed object in the case of a UAF, the model remains agnostic to the specific characteristics of the \textit{vulnerable object}. This key feature allows Page Spray to be independent of the attributes of the \textit{vulnerable object}, making it an orthogonal approach to addressing vulnerabilities.

This characteristic presents a significant advantage of Page Spray when compared to traditional methods such as heap object spraying. The exploitability of these conventional methods is heavily influenced by the level of control that can be exerted over the \textit{vulnerable object}. In contrast, Page Spray enables attackers to reuse the entire model with minimal modifications when targeting different vulnerabilities of the same type. This flexibility and independence from the specific attributes of the \textit{vulnerable object} enhance Page Spray's effectiveness in exploiting a wide range of kernel vulnerabilities.

%% file: sections/5-callsite_model.tex
\section{Root Cause of Page Spray}
\label{sec:rootcause}
In this section, we focus on analyzing the root cause of the Page Spray exploitation technique. Additionally, we provide real-world case analysis to aid in understanding.

\subsection{Existence of Fast Page-level Operations}
Based on the high-level exploitation model shown in Figure~\ref{fig:exploitmodel}, page spraying in the Linux kernel heavily depends on the use of \textit{page-spraying kernel callsites} to achieve Step 4. The primary distinction between page-spraying callsites and regular page-allocation functions lies in the control they offer over data within kernel pages. Page-spraying callsites not only allocate pages but also provide opportunities for attackers to manipulate page-level memory spaces directly (compared to indirect operations like SLUB subsystem), allowing them to place malicious page-level data as desired.



After diving into the kernel codebase and analyzing concrete patterns that allow attackers to control page allocation and its content, we categorize page-spraying callsites into two main types: Copy-Write Calls and Remapping Calls. The design of these two types of calls primarily revolve around enhancing system performance. Specifically, during periods of high memory demand in the system, the smaller non-page-level buffers prove insufficient for promptly storing the substantial data volume. Consequently, the system requires the rapid allocation of larger page-level buffers, leading to direct page-level allocations.

As for the scenario in remapping (zero-copy) callsites. Considering the sensitivity of performance-critical functionality, performing multiple data copies between user space and kernel space incurs significant performance overhead. As a result, the concept of zero-copy design emerged, where the kernel dynamically allocates page-level buffers at runtime and shares these memory regions with the user space as needed in order to reduce the aforementioned overhead and increase data handling capacity.

These designs play irreplaceable roles in ensuring the kernel's performance and reliability. Their presence is essential for maintaining high availability and enhancing overall performance. Consequently, it is impossible to eliminate or reduce them as part of any mitigation strategy within the kernel. In other words, the root cause of Page Spray cannot be eliminated.

\subsection{Copy-Write Call Mode}
\label{call-write-call-model}
To elaborate on this call mode, we can identify two sub-types of copy-write operations. It's essential to note that both of these sub-types of calls can be employed with Page Spray techniques. However, it's crucial to highlight that their low-level design intricacies differ within the Linux Kernel.

\paragraph{Raw Page-Level Buffer.}
In consideration of the kernel's design, several subsystems incorporate a function that initially sets up a buffer and subsequently triggers and writes to this buffer as needed. An example of this can be observed in the code snippet shown in List~\ref{lst:pipewrite}, which is part of the pipe subsystem.

Within the pipe subsystem, a meta-control structure responsible for buffer management is referred to as \code{pipe_buffer}. This structure employs a pointer to page(s) to locate and access the actual buffer when there's a buffer-write request originating from user space. In the \code{pipe_write()} function's operations, after simplification, it's evident that the kernel's intention is to first allocate multiple pages as required. Subsequently, these allocated pages are assigned to the page buffer pointer inside the corresponding \code{pipe_buffer} object. Finally, the function \code{copy_page_from_iter()} is invoked to carry out the specific copy-write operations.

It's worth mentioning that when the \code{pipe_buffer} is generated and initialized, the page buffer is not allocated and assigned immediately in practice. Instead, the kernel defers the allocation of page buffers until a buffer's copy-write request is received and processed. This represents a delay-allocation strategy employed by the kernel for page buffers.

\begin{figure}
\begin{lstlisting}[caption={Copy-Write implementation in Pipe subsystem.},label={lst:pipewrite}]
struct pipe_buffer {
	struct page *page;
	unsigned int offset, len;
	const struct pipe_buf_operations *ops;
        ...
};
static ssize_t
pipe_write(..., struct iov_iter *from) {
for (;;) {
    if (!page) {
        page = alloc_page(GFP_HIGHUSER | __GFP_ACCOUNT);
        ...
    }
    buf->page = page;
    copied = copy_page_from_iter(page, 0, PAGE_SIZE, from);
    }
}
\end{lstlisting}
\vspace{-4ex}
\end{figure}

\paragraph{Non-linear Page-Frags Buffer.}
Within the networking subsystem, the structure known as \code{sk_buff}~\cite{skb} is responsible for representing network packets. To focus solely on relevant details, we concentrate on the buffer aspects and omit unrelated components. In this context, the design of the buffer can be divided into two types: linear buffer and non-linear buffer.

The linear buffer is allocated at the early stage of \code{skb} creation and primarily serves to hold protocol headers and a portion of actual data. Page allocation does not occur at this initial stage. On the other hand, the non-linear buffer comes into play when the linear buffer is unable to provide sufficient memory space for data storage. The organization of the non-linear buffer resembles multiple fragments, with each fragment maintaining a specific page as the data buffer. The kernel employs copy-write operations to transfer data from user space to these continuous page fragments.

In the relevant function \code{packet_snd} (as seen in List~\ref{lst:packetsnd}), an \code{skb} is generated by \code{alloc_skb_with_frags()} in lower level, which is called to allocate page fragments as buffers. Subsequently, the \code{skb_copy_datagram_from_iter()} function manages the copy-write process by invoking \code{copy_page_from_iter()}. When the kernel initiates the copy-write procedure, the page fragments are transformed into real page addresses using the \code{skb_frag_page()} function.

\begin{figure}
\begin{lstlisting}[caption={Copy-Write implementation in Packet Address Family.},label={lst:packetsnd},language=C]
typedef struct bio_vec skb_frag_t;
static int packet_snd(struct socket *sock, struct msghdr *msg, size_t len) {
    ...
    skb = packet_alloc_skb(sk, hlen + tlen, hlen, len, linear, msg->msg_flags & MSG_DONTWAIT, &err);
    ...
    err = skb_copy_datagram_from_iter(skb, offset, &msg->msg_iter, len);
}
\end{lstlisting}
\vspace{-4ex}
\end{figure}

\subsection{Remapping Call Mode}

In this subsection, we discuss another major type of Page Spray calls, remapping calls. To demonstrate the Page Spray's ingenious invoking mode, we also introduce code segments to illustrate its nature and particularized implementation in kernel. We will discuss the copy-write and zero-copy design in Linux Kernel, then explore in depth the model. 

\paragraph{Mmap \& Zero Copy Calls.} Typically, when a user-space application executes a \code{mmap()} call to establish a new mapping in the virtual address space of the calling process, the kernel primarily grants access rights to the virtual address as the return value of this call.

When we delve into the kernel implementation, we found the zero-copy mechanism facilitated by memory mapping opens up opportunities for Page Spray. Where data copy operations occur between a user process and kernel space, the boundary between these two memory address spaces often necessitates checks for operation validity and data reconstruction in the kernel space. Therefore, copy operations that traverse the boundary between user processes and kernel space can raise significant performance concerns, especially in subsystems that prioritize performance, such as the networking subsystem. To address this challenge and improve overall system performance, Linux incorporates a zero-copy design. With zero-copy, data transfer between kernel space and user processes reduces the additional overhead introduced by memory copies.

In this design, user-space processes explicitly set (ring) buffer attributes, and the allocated pages are organized into multiple page-related structures. When a process from user space perform a \code{mmap()} call, its detailed implementation and execution are redirected by the corresponding kernel function tables based on the socket's attributes. For instance, when setting up a packet ring buffer in PF\_PACKET, the \code{mmap()} call is executed by the \code{packet_mmap()} function in the kernel (as seen in List~\ref{lst:packetmmap}). In this implementation, the page-related structure it introduces is page-vector, which can be used to maintain a reference to a valid page and akin to the previously discussed skb page-fragments. The kernel first transforms a page-vector to a raw page reference in Line 8. Subsequently, in the next line, \code{vm_insert_page()} is invoked to insert the target page into the virtual address mapping space of the user space process. After this insertion, the user space process gains the right to achieve direct memory write operations to the page. In other words, at this stage, the related user process becomes capable of spraying data onto pages which allocated by the kernel space.

\begin{figure}
\begin{lstlisting}[caption={Example codes of mapping in packet mmap.}, label={lst:packetmmap}]
static int packet_mmap(..., struct vm_area_struct *vma){
...
for (i = 0; i < rb->pg_vec_len; i++) {
    struct page *page;
    void *kaddr = rb->pg_vec[i].buffer;
    int pg_num;
    for (pg_num = 0; pg_num < rb->pg_vec_pages; pg_num++) {
		page = pgv_to_page(kaddr);
		err = vm_insert_page(vma, start, page);
        ...
        }
    }
}
\end{lstlisting}
\vspace{-4ex}
\end{figure}

The zero-copy mapping mechanism leads to a page of memory explicitly shared between kernel space and user space, allowing bad actors to overcome previous limitations. What sets zero-copy mapping apart from other methods is that it inherently allocates pages for data (ring) buffers and makes them accessible to user space. In this scenario, user space gains flexible control over memory allocation and writing behaviors. For example, user space can explicitly call \code{packet_set_ring()} to specify the number of pages to allocate (as shown in List~\ref{lst:packetsetring}), where the \code{req} variable represents the ring buffer configuration request sent from the user process.

Furthermore, zero-copy mapping introduces kernel virtual address space overlaps between slab pages and zero-copy pages, providing a fundamental support for Page Spray. This means that a page discarded by the kernel heap's slab allocator can be reclaimed through zero-copy operations and remapped to the memory space of the user space process.

\begin{figure}
\begin{lstlisting}[caption={Example codes of page allocation in packet ringbuffer setting.},label={lst:packetsetring}]
static struct pgv *alloc_pg_vec(struct tpacket_req *req, int order){
    unsigned int block_nr = req->tp_block_nr;
    pg_vec = kcalloc(block_nr, sizeof(struct pgv), GFP_KERNEL | __GFP_NOWARN);
    for (i = 0; i < block_nr; i++) {
        pg_vec[i].buffer = alloc_one_pg_vec_page(order);
        ...
    }
}
\end{lstlisting}
\vspace{-4ex}
\end{figure}

%% file: sections/6-static_analysis_model.tex
\section{Static Analysis Model}
\label{sec:static}
To understand the extent to which Page Spray impacts the security of the kernel, it is crucial to comprehend the prevalence and generality of Page Spray in the Linux kernel. While manually identifying all potential call sites is the most straightforward approach to consider, the intricacy of the Linux kernel codebase makes manual selection relatively impractical and inadequate for supporting a large-scale comprehensive study.

To tackle this challenge and offer a systematic comprehension, we introduce our static analysis model designed to scrutinize the Linux kernel codebase. Furthermore, leveraging this model, we develop an automated static analyzer tool and apply it to the kernel codebase.

In this section, we introduce our static analysis model aimed at pinpointing the crucial page-spraying call sites within the Linux Kernel through a combination of control-flow analysis and data-flow analysis.

\subsection{Static Analysis Methodology}
\paragraph{Copy-Write Callsites.} To identify Copy-Write page-spraying callsites, our approach involves the analysis from root interfaces, which we categorize into two distinct types: allocation and copy root interfaces. 

Allocation root interfaces primarily serve the purpose of executing direct page allocation or functioning as wrapper functions within specific subsystems to accomplish such allocations. It is worth noting that, although some page allocations occur in the kernel heap slab allocator, we do not consider those as the qualified callsites for Page Spray. The distinction here arises from the granularity of operations within the slab allocator, which operates at the object-level rather than the page-level, and in Page Spray, we consider direct page-level allocation.

On the other hand, copy interfaces are mainly associated with copy-write functions in the kernel. These interfaces are capable of transferring of data from user space to the kernel's page buffers. Through these operations, user processes gain the capability to manipulate data at the page level. Because our data copy are based on the "benign" operations, such as copying a data to kernel space networking buffer and so on, it natively will not trigger any defense or mitigation like SMAP which is designed for enhancing islation between user space and kernel space.

To systematically identify callsites, we build a global-level callgraph in the kernel to capture inter-function control flow information. Each function call is represented as a node within this graph. We initiate our search by tracing backward from two root interface call nodes: one pertaining to allocation and the other to copy operations. As we traceback paths from root interfaces, we identify nodes where the paths intersect. These nodes are marked as potential callsites at the control flow level.

Then, to enhance the accuracy, we employ fine-grained data flow analysis. This entails conducting data flow analysis inside individual nodes. Our data flow analysis is guided by two specific instruction points. We ensure that the allocation instruction precedes the copy-write operation. Subsequently, we proceed to a search process from the allocation points to the copy-write points, with the aim of ascertaining whether the corresponding data transmission occurs between them. Specifically, if the allocation operation allocates pages and assigns them to a certain position in a control structure, and the subsequently corresponding copy-write operation writes data to that same position of the structure, we recognize this as a valid callsite.

\begin{figure}[t]
\centering
\includegraphics[width=\columnwidth]{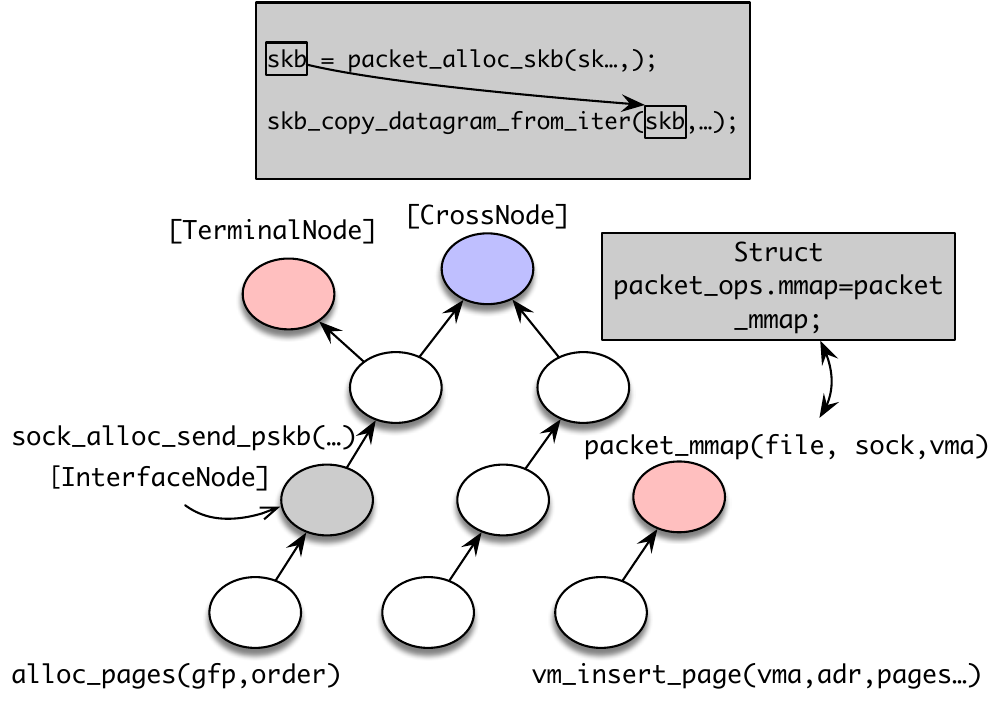}
\caption{Page Spray Invocation Model. Each node indicates a specific function in the overall CallGraph. \code{skb} is the structure which is binded with page frags. }
\label{fig:callsitemode}
\vspace{-10px}
\end{figure}

\paragraph{Remapping Callsites. }On the contrary, when examining zero-copy calls, these are tightly linked to the function tables of specific subsystems, such as the \code{packet_ops}, in various protocol families. Therefore, in our analysis, we start by traversing the structures of these function tables at an early stage and gather pertinent information regarding function pointer members. Of particular interest is the mmap-related function pointer, which directs the mmap-execution-flow to the actual implementation in the kernel. Much like in our previous analysis, zero-copy mmap calls also contain memory remapping root interfaces. For instance, \code{vm_insert_page()} is responsible for inserting a single page into a user's virtual memory area, and \code{remap_pfn_range()} facilitates the remapping of kernel memory for sharing with user space processes. 

In the zero-copy mapping model analysis, these root interfaces are also treated as nodes at the control flow level. During our analysis, we trace back from these nodes. When we encounter a specific \code{*_mmap} function that aligns with the mmap function within a function pointer found in our collected function tables, we preliminarily consider it as a zero-copy remapping callsite. Furthermore, at the data flow level, we apply forward dataflow analysis and backward dataflow analysis. We backtrack from the allocation root interfaces to ascertain whether the same control structure is employed both in allocation and remapping. If we identify a structure that connects allocation with remapping, we recognize the callsite as a valid zero-copy remapping type page-spraying callsite.

\paragraph{Optimization.} As illustrated in Figure~\ref{fig:callsitemode}, the interface nodes represent the starting points for one-time search. While there are some raw nodes positioned beneath these interface nodes, we have opted to exclude these nodes to simplify the backtrace process. Like the pruning operation, the terminal nodes point to functions that we've encountered during the analysis but are not intended to be traced back to their higher parent function nodes. The introduction of this is necessary for several reasons. Firstly, in the slab allocator, there are functions that rely on the page allocator but do not directly trigger page allocation, like \code{kmalloc()}. Although indirect allocation may occur when a new slab allocation is executed, it does not align with the scenarios capable of triggering Page Spray. Secondly, the backtrace operations may have reached their maximum limit, so they terminate as we expect. Lastly, there are sections of code within the kernel that are exclusively available on specific architectures or hardware. In such cases, we dissociate the results from these code segments to ensure accuracy and relevance.

It is crucial to note that during the process of remapping invocation, there might be instances where there is no "Crossing Node." This occurs because the remapping \code{mmap()} function often follows a separate execution path from allocation. To fit this situation, we set the related "Terminal Node" for it. For example, when a \code{mmap()} call is made in the user space for a packet page buffer, it is typically executed by \code{packet_mmap()} in kernel. Therefore, \code{packet_mmap()} is considered a "Terminal Node". If the analyzer encounters this point during backtrace, it immediately stops and proceeds to analyze allocation interfaces to determine if the data flow between allocation and remapping aligns appropriately. As for allocation, we backtrace from the direct page allocation points, like \code{__get_free_page()}. Considering the multiple reference and the depth of invocation chains for these lower direct page allocation calls, when we find the tracing enter the related subsystem which is the same as previous mmap invocation, we will reset the root interface into the function in corresponding subsystem, and searching the control structure in this given subsystem. If the same control structure is found, we will compare with the control structure of the mmap operations.

\paragraph{Design \& Implementation.} Due to the page limit, we illustrate the implementation information and have a discussion. in ~\ref{sec:appendix2}.

\subsection{Dynamic Test and Analysis}
\label{sec:syzkaller-test}
To ensure the effectiveness of the callsite analysis, we deployed the state-of-the-art and most widely used kernel fuzzer, Syzkaller~\cite{syzkaller}. In order to enable Syzkaller to report the corresponding callsites' reachability, we insert \code{WARN_ON(1)} as the panic function in each callsite. The Syzkaller will generate syscall sequences to reach and trigger the callsite, once it reaches, the kernel will experience an error by warning messages and stack dump information. In the following steps, Syzkaller will try to reproduce the related panic then output a minimized input. It should be emphasized that Syzkaller is performed based on syscall templates to generate the fuzzing input and  dynamically test the kernel. For some kernel modules and logics, the templates have not yet been supported well to fit the corresponding part by Syzkaller. In these kinds of situation, we confirm the callsite manually with our crafted trigger programs. During our dynamic test, manually check only happens for \code{tap_skb_alloc} callsite. 

For configuration of Syzkaller, we set 8 processes with qemu-based instance in bare-mental machine. Every instance we set 2 CPUs with 2G memory. 

\subsection{Analysis Result}
\label{sec:analysis-result}
As shown in Table~\ref{tab:callsites}, by our analyzer, we have identified 21 page-spraying callsites within the Linux kernel source code. These results illustrate that the page-spraying callsites are distributed across several crucial subsystems in the Linux Kernel, including Networking, Pipe, io\_uring and so on. Notably, the networking subsystem in Linux Kernel exhibits a higher frequency of these page-spraying callsites. Upon closer examination, we have observed that this frequency can be attributed to the data transmission model employed in the networking subsystem.

When transmitting data or messages via various protocols in Linux Kernel, non-linear buffers may be allocated and utilized for data storage. The commonly used kernel function \code{alloc_skb_with_frags()} is responsible for allocating kernel pages from the page allocator to store data received from user space. Furthermore, we have identified memory remapping-related callsites within the networking/pipe subsystem and the \textit{io\_uring} subsystem. These callsites, along with their respective subsystems, are designed to share memory spaces with user space during the early stages of constructing (ring) buffers. This approach aims to reduce the overhead associated with cross-space data copying and writing, ultimately improving system performance, as we discussed previously.

Moreover, we conducted a dynamic analysis of each callsite identified by the static analyzer as validation. We perform a Syzkaller test in kernel level for all callsites. For our results, Syzkaller automatically triggers 20 of 21 in our callsites. For callsite \code{tap_alloc_skb()} which is actually inlined in \code{tap_get_user()}, we find that Syzkaller has trouble triggering it. After investigating the information from syzbot-assets\cite{syzbot-assets} of Syzkaller, we learn that Syzkaller has a low coverage in tap.c file. Due to this reason, only for this callsite, we use our crafted program based on the testcase from Linux v6.1 selftests\cite{linux-selftest} to trigger it manually.

Beyond the current findings, our static analyzer can be utilized for continuous integration. As the Linux kernel continues to evolve, incorporating new functionalities and features, we can re-employ the analyzer to detect recently added page-spraying callsites.

\begin{table}[t]
    \tabcolsep=4pt
    \centering
        \begin{tabular}{llllllllc}
         \toprule
            \normalsize{\textbf{Callsite}} & 
            \normalsize{\textbf{Usability}} & 
            \normalsize{\textbf{Syscall}} &
            \\
         \midrule
         packet\_set\_ring &  \fullcirc & setsockopt\\
         packet\_snd &  \fullcirc & sendmsg\\
         packet\_mmap &  \fullcirc & mmap\\
         rds\_message\_copy\_from\_user &  \fullcirc & sendmsg\\
         unix\_dgram\_sendmsg & \halfcirc$\dagger$ & sendmsg\\
         unix\_stream\_sendmsg &  \halfcirc$\dagger$ & sendmsg\\
         netlink\_sendmsg &  \halfcirc\ding{60} & sendmsg\\
         tcp\_send\_rcvq(inet6) &  \fullcirc & sendto\\
         tcp\_send\_rcvq & \fullcirc & sendto\\
         tun\_build\_skb &  \halfcirc\ding{63} & write\\
         tun\_alloc\_skb & \halfcirc\ding{63} & write\\
         tap\_alloc\_skb & \halfcirc\ding{63} & write\\
         pipe\_write  &\fullcirc & write &\\
         fuse\_do\_ioctl  & \halfcirc$\dagger$ & ioctl\\
         io\_uring\_mmap  & \fullcirc & mmap\\
         array\_map\_mmap  & \halfcirc$\dagger$ & mmap\\
         ringbuf\_map\_mmap  & \halfcirc$\dagger$ & mmap\\
         aead\_sendmsg  & \fullcirc & sendmsg &\\
         skcipher\_sendmsg  & \fullcirc & sendmsg &\\
         mptcp\_sendmsg  & \fullcirc & sendmsg\\
         xsk\_mmap  & \halfcirc\ding{62} & mmap\\
       \bottomrule
        \end{tabular}
        \caption{The page-spraying callsites analysis results for Page Spray. \fullcirc:  represents callsites that can be used without extra restriction. \halfcirc:  represents callsites that are capable of exploiting with limitations. $\dagger$ means the usability is restricted by allocation size. \ding{63} indicates the requirements of device access. \ding{61} represents callsite requires XDP operations. \ding{60} means the memory area restriction.}
        \vspace{-2ex}
        \label{tab:callsites}
\end{table}

%% file: sections/7-evaluation.tex
\begin{table}[t]\small
    \centering
    \begin{tabular}{l|c|c|cc}
        \toprule
        \textbf{CVE-ID} &  \textbf{Type} & \textbf{Object Spray} & \textbf{Page Spray}\\
        \midrule
        \rowcolor{gray!30} 
        CVE-2016-4557          & UAF & \ding{52} & \ding{52} \\
        CVE-2016-8655       & UAF & \ding{52} & \ding{52}  \\
        \rowcolor{gray!30}
        CVE-2017-10661       & UAF & \ding{52} & \ding{52}\\
        CVE-2017-11176          & UAF & \ding{52} & \ding{52} \\
        \rowcolor{gray!30}
        CVE-2017-15649          & UAF & \ding{52} & \ding{52}\\
        CVE-2018-6555    & UAF & \ding{52} & \ding{52} \\
        \rowcolor{gray!30}
        CVE-2016-0728    & OOB & \ding{52} & \ding{52} \\
        CVE-2021-22555      & OOB & \ding{52} & \ding{52}\\
        \rowcolor{gray!30}
        CVE-2022-2588       & DF & \ding{52} & \ding{52}\\
        CVE-2017-6074  & DF & \ding{52} & \ding{52}\\
        \rowcolor{gray!30}
        CVE-2017-8890 & DF & \ding{52} & \ding{52} \\
        CVE-2022-29581 $\dagger$ & UAF & \ding{52} & \ding{52} \\
            \rowcolor{gray!30}
        CVE-2016-10150 & UAF & \ding{52} & \ding{55} \\
        
        CVE-2022-20409 \ding{72} & UAF & \ding{52} & \ding{52} \\
        \rowcolor{gray!30}
        CVE-2022-2585 $\dagger$ & UAF & \ding{55} & \ding{52} \\
        \bottomrule
    \end{tabular}
    \caption{Exploitability demonstrated on real-world vulnerabilities. The symbol \ding{72} represents we exploit this vulnerability on Mobile Device. $\dagger$ means that the exploitation in this case includes 
    cross-cache.}
    \vspace{-5ex}. 
    \label{tab:exploitability}
\end{table}

\begin{table*}[t]\small
\centering
\vspace{0px}
\begin{tabular}{lllllll}

\toprule
\textbf{Type} & \textbf{CVE} & \textbf{Slab-Cache}  & \textbf{Single-Thread Spray} & \textbf{Multi-Process Spray} & \textbf{Page Spray} & \textbf{Subtypes} \\
\midrule
\multicolumn{7}{c}{\textbf{In IDLE State}} \\
\midrule
\rowcolor{gray!30} 
UAF & CVE-2016-4557 $\dagger$ & Kmalloc-256 & 100\% & 100\% & 100\% & eBPF \\
UAF & CVE-2016-8655 $\dagger$ & Kmalloc-2048 & 99.4\% & 99.3\% & 100\% & Race\\
\rowcolor{gray!30} 
UAF & CVE-2017-10661 $\dagger$ & Kmalloc-256 & 41.4\% & 64.1\% & 99.8\% & Race\\
UAF & CVE-2017-11176 $\dagger$ & Kmalloc-2048 & 99.4\% & 99.8\% & 99.7\% &  Normal\\
\rowcolor{gray!30} 
UAF & CVE-2017-15649 $\dagger$ & Kmalloc-4096 & 61.4\% & 99.4\% & 97.9\% & Race\\
UAF & CVE-2018-6555 & Kmalloc-96 & 98.9\% & 100\% & 87.7\% &  Normal \\
\rowcolor{gray!30} 
OOB & CVE-2016-0728 $\dagger$ & Kmalloc-256 & 91.3\% & 99.8\% & 99.3\% & Race \\
OOB & CVE-2021-22555 & Kmalloc-1024 & 77.3\% & 46.0\% & 61.2\% &  Normal \\
\rowcolor{gray!30} 
DF & CVE-2017-8890 $\dagger$ & Kmalloc-64 & 74.3\% & 94.6\% & 94.4\% &  Normal\\
DF & CVE-2022-2588 $\dagger$ & Kmalloc-256 & 87.3\% & 10.6\% & 91.4\% & Normal\\
\midrule
\multicolumn{7}{c}{\textbf{In BUSY State (stress-ng)}} \\
\midrule
\rowcolor{gray!30} 
UAF & CVE-2016-4557 & Kmalloc-256 & 75.6\% & 97.4\% & 84.4\% &  eBPF \\
UAF & CVE-2016-8655 $\dagger$  & Kmalloc-2048 & 64.3\% & 58.1\% & 61.5\% & Race\\
\rowcolor{gray!30} 
UAF & CVE-2017-10661 $\dagger$  & Kmalloc-256 & 28.6\% & 78.3\% & 98.1\% & Race\\
UAF & CVE-2017-11176 & Kmalloc-2048 & 79.8\% & 94.4\% & 63.7\% & Normal\\
\rowcolor{gray!30} 
UAF & CVE-2017-15649 $\dagger$  & Kmalloc-4096 & 38.1\% & 98.8\% & 99.2\% & Race\\
UAF & CVE-2018-6555 $\dagger$  & Kmalloc-96 & 92.0\% & 98.1\% & 90.7\% & Normal\\
\rowcolor{gray!30} 
OOB & CVE-2016-0728 & Kmalloc-256 & 40.4\% & 99.9\% & 87.3\% & Race \\
OOB & CVE-2021-22555 & Kmalloc-1024 & 71.8\% & 39.4\% & 43.4\% & Normal \\ 
\rowcolor{gray!30} 
DF & CVE-2017-8890 $\dagger$  & Kmalloc-64 & 18.7\% & 27.8\% & 49.0\% & Normal\\
DF & CVE-2022-2588 $\dagger$   & Kmalloc-256 & 50.9\% & 19.0\% & 54.0\% & Normal\\
\bottomrule
\end{tabular}
\caption{Exploit Stability Evaluation of Page Spray under Idle State and Busy State generated by \code{stress-ng}, compared to Single-Thread Spray and Multi-Process Spray. The $\dagger$ symbols represent cases Page Spray achieves better or comparable results as the traditional approaches. }
\label{tab:stability}
\vspace{-10px}
\end{table*}

\section{Effectiveness Evaluation}
\label{sec:evaluation}
In this section, we will discuss the design and results of several experiments conducted to measure the effectiveness of the Page Spray technique in the context of various real-world kernel vulnerabilities. To gain a comprehensive understanding, we categorize "effectiveness" into two distinct attributes: "exploitability" and "stability". The results of exploitability evaluation are shown in Table~\ref{tab:exploitability}, and the stability evaluation results are in Table ~\ref{tab:stability}.

\subsection{Experiment Design \& Methodology}
We have executed a series of experiments to investigate the practicality of employing the Page Spray technique in the exploitation of real-world kernel vulnerabilities. These experiments underscore the inherent complexity of real-world vulnerabilities exploitation, which can be attributed to several key factors.

Firstly, the subsystem within the kernel where vulnerabilities manifest can significantly influence the contexts of the exploitation process. An effective exploitation method should ideally be versatile enough to overcome or mitigate the influence of these context-specific variations, with as few adjustments as possible. Secondly, the nature of the vulnerable object plays an imperative role in shaping the exploitation strategy. Objects of different sizes and behaviors require distinct approaches to achieve successful exploitation. Lastly, the patterns exhibited by vulnerabilities, such as Use-After-Free, Double-Free, Invalid-Free, and Out-of-Bound, introduce further complexity. Attackers must select appropriate exploitation methods that align with these specific patterns in order to effectively exploit the vulnerabilities.

Based on the outcomes of these experiments, the effectiveness of Page Spray in the context of real-world vulnerabilities is evaluated. Additionally, a comprehensive analysis is conducted to evaluate how proficiently Page Spray performs when applied to the exploitation of these vulnerabilities.

In our evaluation, we systematically assess a selection of real-world Linux kernel CVEs, adhering to the methodology outlined in K(H)eaps~\cite{kheaps}. Furthermore, we introduce several up-to-date CVEs and real-world vulnerabilities into our assessment. Our criteria for CVE selection focus on kernel heap data-related corruptions that align with the principles established in K(H)eaps. These CVEs are rigorously verified for their reproducibility within the Linux kernel and their capacity to induce kernel crashes. 

Our collection of CVEs encompasses a total of 15 real-world Linux kernel CVEs, as presented in Table~\ref{tab:exploitability}. Our collection incorporates a set of 9 native Use-After-Free bugs, 3 Double-Free bugs, and 2 Out-Of-Bound bugs. This diverse set of vulnerabilities includes various subtypes such as race conditions, eBPF-related issues, and cross-cache attacks~\cite{cross-cache1,cross-cache2}. 

\paragraph{Exploitability Evaluation. } 
To evaluate the exploitability of Page Spray, we adopt a straightforward approach. We create a Page Spray proof-of-concept (PoC) for each of the CVEs within our dataset. Subsequently, we execute these PoCs to ascertain how many of the CVEs can be successfully exploited using Page Spray with a reasonable success rate. In instances where our initial PoC fails, we conduct an in-depth examination to identify the underlying reasons and try to adapt the PoC to the specific circumstances. In cases where no modifications yield successful exploitation, we draw the conclusion that Page Spray is incapable of exploiting that particular vulnerability and provide a detailed explanation for this outcome.

\paragraph{Stability Evaluation. } 
Our stability evaluation experiments involve running exploits for a subset of the CVEs that are exploitable by Page Spray from the exploitability evaluation for a 1000 iterations under two distinct system operational states: the idle working state and the busy working state. The idle state represents a moderate kernel heap activity environment on the machine. This state mirrors the conditions that one would typically encounter on a standard local user's machine. Conversely, the busy state aims to simulate a more extreme operational setting where many intensive applications and stressors are running concurrently, providing a comprehensive context to limit test the capability of the Page Spray technique. We simulate this busy working state by utilizing a benchmark program called \code{stress-ng}~\cite{stressng}. Due to the space limit, we describe the detail of the environment used for evaluation in Appendix \S\ref{sec:appendix1}.

Under these varying working conditions, we conduct a comparative analysis involving several practical state-of-the-art kernel heap exploitation techniques, including Page Spray, Single-Thread Heap Spray, and Multi-Process Heap Spray, as analyzed in K(H)eaps. This analysis serves to reveal the potential advantages or disadvantages of the Page Spray technique in the context of real-world environments.

The exploits are developed by first collecting publicly available and approved PoCs for each CVE. Then, we modify them into corresponding variants (Page Spray, Single-Threaded/Multi-Process Heap Spray) following the method we propose in Section \S\ref{sec:expmodel}. To ensure fairness, the only difference between each exploitation variant is the spraying technique itself, based on the same setups. Different types of vulnerabilities have slight differences compared to our proposed method in section 4.1, as discussed in sections 4.2 and 4.3. A step-by-step detail of one such exploits can be found in Section \S\ref{sec:refurbish} and \S\ref{sec:appendix3}.

\subsection{Experiment Results}

\paragraph{Exploitability Evaluation. }
The outcome of our experiment is presented in Table~\ref{tab:exploitability}. Page Spray demonstrated successful exploitation in 14 out of 15 vulnerabilities within our dataset. These successful exploits encompassed all three vulnerability types: UAF, OOB, and Double Free. Notably, Page Spray also effectively exploited a vulnerability on a mobile device (CVE-2022-20409) and executed cross-cache attacks (CVE-2022-2585, CVE-2022-29581). 

Specifically, in the case of CVE-2022-2585, which is a relatively recent kernel heap vulnerability, traditional object spray method is unable to successfully exploit it due to the lack of precise control over the entirety of a vulnerable object and insufficient kernel information leakage. In contrast, Page Spray excels in this scenario. It effectively exploits CVE-2022-2585, and we'll delve into this in more detail in Section \S\ref{sec:refurbish}, where we explore how Page Spray shines in addressing recent real-world vulnerabilities.
 
On the other hand, in the case where Page Spray failed, we conducted a deep analysis to find out the root cause. In CVE-2016-10150, the vulnerability is triggered in KVM kernel's module. The UAF's allocation and free happen during a single invocation to \code{ioctl()}. This situation introduces a challenging race condition with a very narrow time window. This is a corner case for Page Spray, since as demonstrated in \S\ref{sec:expmodel}, the page layout manipulation and page spraying must occur after the vulnerable object is allocated and before it is freed to effectively exploit the vulnerability. However, the stringent time constraints imposed by this case make it exceedingly difficult to set up the necessary conditions for Page Spray to be successful.

\paragraph{Stability Evaluation. }
As evident from our observations of the experiment result in Table~\ref{tab:stability}, in an idle system environment, the Page Spray technique demonstrates remarkable effectiveness, surpassing both the Single-Thread Heap Spray and Multi-Process Heap Spray methods. It consistently achieves an impressively high success rate, often approaching 100\%, particularly in scenarios involving UAF vulnerabilities. This robust performance is noteworthy because real-world kernel heap exploits commonly occur on systems with moderate workloads. For instance, when jailbreaking mobile devices, one of the common applications of kernel exploitation, the jailbreaker typically has complete control over the device's workload and can reduce it to a minimum. This level of workload can also be expected on a standard local user's machine, which is a common target by attackers for exploitation. Furthermore, in attacks on busy remote servers like web servers or cloud servers, attackers can monitor the server's workload and strategically time their exploits during periods of lower activity, such as late at night.

In contrast, when operating within extremely stressed environments, it is expected that the success rates of Page Spray, as well as the other compared techniques, would experience a decrease. Nevertheless, Page Spray maintains a superior success rate in comparison to the Single-Thread Heap Spray method, and it delivers results comparable to those achieved with Multi-Process Heap Spray. Notably, in specific circumstances, Page Spray even performs on par or significantly better than Multi-Process Heap Spray, as exemplified by CVE-2017-10661, CVE-2017-15649, CVE-2018-6555, CVE-2017-8890, and CVE-2022-2588. The forthcoming subsection will delve into the conditions under which Page Spray excels and elucidate the rationale behind its performance.

In summary, the outcomes of both evaluations illustrate that Page Spray is a robust and effective approach that can serve as a valuable complement to conventional object spray methods. When used in the correct scenarios, these techniques cover each other's limitations, leading to a broader range of vulnerabilities that can be exploited and ultimately enhancing the stability of successful exploits.

\subsection{Result Discussion}
\paragraph{Root Cause of Instability. }
In order to explain and have a better understanding of the evaluation results and subsequent comparisons, it is imperative to first understand the key factor that influences the success rate of exploitation techniques and how they differentiate between Page Spray and conventional object spray approach. In the context of object spraying, in order to reclaim the vulnerable object’s slot, the object must be in the active slab. Consequently, the exploit program has to race against the unpredictable and random allocations coming from the busy kernel with the goal to reclaim that specific slot for the desired object. 

In contrast, page spraying can mitigate this issue, by spraying the target cache with spray objects until the current slab is full, and another slab is allocated as the active slab, all before freeing the vulnerable object. In this case, all the unpredictable allocations from the kernel are directed into the active slab, while the freed vulnerable object resides in another slab, remaining untouched. This is possible because in this technique, the goal is not to reclaim the individual object, but rather to target the entire page for reclamation. However, it is worth noting that the stability concern in the context of page spraying arises during the actual spraying process itself. The key requirement for the pages to be recycled to the allocator is that all objects within the entire slab containing the vulnerable object must be freed. Consequently, there must be no unpredicted allocation within the same slab, as this would impede this requirement.


This characteristic is an advantage for Page Spray under certain conditions. For instance, in the case of CVE-2017-10661, Page Spray shows an remarkable advantage over traditional approach,  boasting a success rate of 98.1\% compared to the traditional method's 78.3\%. A similar trend is observed in CVE-2017-8890, where Page Spray achieves a success rate of 49.0\% compared to the traditional approach's 27.8\%. We specifically examined these cases and it became evident that they have one important thing in common: the vulnerable objects are freed in RCU~\cite{RCU}. This means that the vulnerable objects undergo a grace period before they are returned to the allocator. Importantly, the duration of this grace period is not visible to the exploit program. Therefore, the common solution is to let the process sleep for a few seconds before attempting to reclaim the vulnerable object's slot. As mentioned above, the race to reclaim this slot is the most critical source of instability for the traditional heap spraying approach. Therefore, the longer the process remains in the sleep state after the object has been freed, the lower the likelihood of successfully reclaiming it, thus explaining the low success rates observed. Page Spray does not suffer from this problem as the reclamation stage is not the critical source of its instability.

%% file: sections/8-refurbish.tex
\section{Case Study: Refurbish Intractable Exploit}
\label{sec:refurbish}


In this section, we demonstrate how Page Spray make improvements to certain hard-to-exploit vulnerability case, and enhance the exploitability. To achieve this, Page Spray employs two novel approaches, kernel information leakage by new channel, and halting the CPU execution to improve exploitability. We successfully apply Page Spray into a real world zero-day bug (CVE-2022-2585~\cite{CVE-2022-2585}) and achieve privilege escalation. 

\paragraph{Vulnerability Background:} The vulnerability in question resides within the Posix CPU Timer Subsystems of the Linux Kernel. It pertains to a \code{CLOCK_THREAD_CPUTIME_ID} timer, which is utilized to measure the amount of CPU time consumed by a thread. When the function \code{timer_settime()} is invoked, the timer is marked as "armed." If there exists a user-defined time interval, once that interval elapses, the timer is triggered. To provide further context, Linux manages a linked list of timer-associated structures, \code{struct k_itimer}, within \code{struct posix_cputimers}. When a timer interrupt is raised, the kernel checks whether a specific thread has consumed a sufficient amount of CPU time. If the conditions are met, the timer expires and is placed into a firing linked list. Subsequently, all timers within this firing linked list are utilized to trigger the \code{posix_timer_event}, leading to the issuance of signals to the program.

This UAF vulnerability manifests in scenarios where a thread establishes a thread CPU timer and then invokes \code{execve()}. When \code{execve()} is called, a clean-up operation is executed on behalf of the process, which involves freeing all timers associated with that process. However, the crucial oversight lies in the failure to remove the reference to the timer within the \code{struct posix_cputimers}. In essence, if the timer was already armed before the \code{execve()} operation, the kernel proceeds to free the timer while still maintaining a reference to it within the doubly linked timer list. Consequently, when the designated time arrives, the kernel traverses the linked list, locates the timer, adds it to the firing linked list, and attempts to trigger the timer, thereby leading to a UAF vulnerability.

\paragraph{Kernel Information Leakage:} In this scenario, we initiate a timer UAF vulnerability within a process and prepare the page vectors for subsequent Page Spray exploitation using \code{AF_PACKET}. Following these preparations, we create a child process directly through a \code{fork()} operation. Within the child process, we monitor the release of the slab page, which eventually returns to the page allocator. The parent process triggers the freeing operation. After the page is freed, we allocate and remap the vulnerable timer page to user space. This remapping is achieved through memory remapping operations, as illustrated in Listing~\ref{lst:packetmmap}, performed by the \code{packet_mmap()}. At this stage, the user gains access to the timer's page; however, it's important to note that the kernel still maintains a reference to this timer. Consequently, any modifications made to the timer's memory data will be immediately reflected in the user space, and vice versa.

In the next step, we initiate an exit process operation. Specifically, the "timer queue" utilized here is, in fact, an RBTree rather than a traditional queue structure. During the execution of the exit operation, the kernel invokes \code{timerqueue_del()} to remove the timer from the RBTree and label it as a "dangling node." This marking is achieved through \code{node->__rb_parent_color = node (RB_CLEAR_NODE)}, indicating that the operation associated with process termination adds a kernel heap pointer that points back to the process's own memory region. Consequently, we can read the remapped memory region from user space, allowing us to easily obtain a kernel heap address.

Finally, we release the sprayed pages back to the page allocator and proceed to spray \code{msg_msg} objects in order to reclaim the pages. At this stage, the kernel heap address we previously leaked serves as the address of a \code{msg_msg} object.

To summarize the previous paragraphs, Page Spray during exploitation introduces a novel channel for leaking kernel information. This is achieved by remapping kernel-allocated pages directly to user space. In essence, this means that by sharing pages with kernel space, users can efficiently monitor and access certain kernel data. This approach eliminates the need to copy data from the kernel space to the user space, thereby simplifying the exploitation process and improving its effectiveness.

\paragraph{Trapping CPU Execution.} After successfully leaking kernel information and running on CPU0, we can trigger the vulnerability once more and reclaim the freed pages using Page Spray. However, it's important to note that, at this stage, we modify the payload used for Page Spray to trap CPU0. This involves manipulating the \code{k_itimer} structure in the payload with the following key modifications: (1) setting the \code{it_requeue_pending} field to a large value, (2) faking the \code{sigq} field to point to the leaked \code{msg_msg} address minus 0x20, (3) faking the \code{cpu_timer}'s \code{head} field within the related \code{k_itimer}, to reference the kernel heap address we previously leaked. These modifications are made to trap the execution of CPU0, allowing us to further control its behavior.

When the timer is triggered, we initiate a search within the previously remapped memory region in user space, with the objective of identifying the specific timer. This identification is achieved by examining the \code{firing} field associated with the relevant \code{k_itimer}. Importantly, following this identification, a modification made to the \code{head} field results in the entrapment of CPU0 within an infinite loop during RBTree operations. As a consequence, we switch execution to CPU1 to carry out the subsequent steps. 

This enforced halt serves two minor yet noteworthy purposes. Firstly, it affords us the necessary time to mitigate potential locking issues by resetting the corresponding member within the \code{k_itimer} structure. Secondly, it ensures that the modified timer is not immediately reintroduced to the firing list, thus preventing any potential corruption of the kernel state. Instead, the timer is scheduled to trigger after a 2-second delay. Once these adjustments are completed, we can return to CPU0 and resume execution. With the manipulation of the \code{sigq} and \code{it_requeue_pending} fields, subsequent write operations effectively overwrite critical fields within the \code{msg_msg} structure, resulting in the leakage of kernel base information. 



Page Spray, as evident in this context, possesses a remarkable capability for precise page-level control. This unique ability empowers us to effectively suspend or trap CPU execution, making it exploitable. Without Page Spray, at this time, the kernel's internal state would likely undergo corruption, making exploitation exceptionally challenging or even unattainable.

%% file: sections/9-defense.tex
\section{\vspace{-3pt}Mitigation Discussion}
\subsection{\vspace{-3pt}General Principles}
\label{sec:generalprinciples}
As we have extensively explained in \S\ref{sec:expmodel}, Page Spray primarily involves reusing pages at the page level. These pages are recycled back to page allocators while slabs are discarded, and then the slab pages are reclaimed from page allocators and overwritten through page-level write/copy operations. To address this issue, the key is to prevent the overlap happening between the pages used for slab objects and the pages used for page-level data buffers. In simpler terms, the idea is to introduce a mechanism that isolates or divides these two sets of pages, ensuring that the same pages are not used both in the SLUB system and for direct page allocation.

\subsection{Rethinking Memory Reuse}
\label{sec:securepagereuse}
To further clarify our discussion on mitigation approaches, we illustrate two real-world examples. In one example, the lightweight solution goes to use GFP flag to separate memory allocation area, while in the other, modifying the slub system within the Linux Kernel.

\paragraph{Our Lightweight Mitigation.} As we discussed earlier in \S\ref{sec:gfpflags}, different GFP flags can influence the allocation zones in the Linux Kernel. One straightforward approach involves adjusting the GFP flags at the points where Page Spray is invoked, directing the allocation to a different memory area that does not overlap with the previous slab memory. Although some other GFP flags can be leveraged to achieve the isolation, to select appropriate flags, it's essential to understand the allocation fallback mechanism. In scenarios where the memory areas best suited for the allocation flags are unavailable for upper-level components, the fallback mechanism is triggered to allocate in the next available zones. At the end of this fallback sequence, the DMA region cannot further fall back to another memory area.

Given that no specific ZONE is initially designated as a reserved isolation area for against Page Spray, we make adjustments to the allocation attributes within the Page Spray callsites in the Linux Kernel. We attach \code{GFP_DMA} to the Page Spray allocation points in kernel, which finally reclaims pages from DMA region, instead of general region. This redirection of allocation operations effectively separates Page Spray allocation from the freed slub pages and eliminates the overlapping. It's important to emphasize that our aim here is to validate the feasibility of this principle, and it's not intended for application in a production environment.

\paragraph{Slab Virtual.} Slab Virtual, a mitigation technique designed by Google and currently maintained outside the Linux Source Code Tree, addresses concerns related to object reuse and cross-cache attacks. This approach extends prior work, such as AUTOSLAB~\cite{autoslab} by Grsecurity and PartitionAlloc~\cite{partitionalloc} by Chromium, by taking a more comprehensive approach. Originally, AUTOSLAB is not enough to mitigate Page Spray, since it only isolates objects at object level, page reuse can still happen if the attacker chooses the same type of object for \textit{padding objects} and \textit{vulnerable object}. 

Conversely, Virtual-Slab operates by creating a new, dedicated virtual memory region (previously an unused hole in the address space). In this region, slab objects are allocated, isolating them from the regular memory allocation process. Additionally, adjustments are made to functions like \code{virt\_to\_phys()} to ensure that the mapping between virtual and physical slab memory becomes permanent. This step is crucial because it prevents the possibility of evading UAF attacks in the physical address space. By combining these measures, Virtual slab significantly reduces the potential for overlap between Page Spray allocations and freed slab pages. However, it introduces incompatibility (e.g. compilation and/or runtime conflicts with common defenses such as KFENCE and KASAN), and noticeable overhead compared to our direct mitigation. Furthermore, it is important to note that as of now, Virtual slab has not been integrated into the Linux Mainline, and therefore the current Linux kernel can still be exploited with Page Spray.

\paragraph{Macro-Performance Overhead.} To assess the macro-performance impact of the aforementioned patches, we implemented both of them in Linux version 6.1. We conducted our experiments on a bare-metal machine equipped with 32GB of memory, a 200GB SSD storage drive, and a 4-core 11th Gen Intel i7 processor running at 2.90GHz. The performance evaluations were executed using Phoronix-Benchmarks~\cite{phoronixbenchmark}. For each specific benchmark case, we ran the tests five times and calculated the average results. The outcomes of our measurements are summarized in Table~\ref{tab:phoronixbench}. Notably, the introduction of Virtual SLUB seems to result in an approximate 4\% macro-overhead in these benchmarks. In contrast, the Hardened patch introduces minimal and practically negligible performance overhead.

\begin{table}[t]\small
    \centering
    \begin{tabular}{lccc}
        \toprule
        \textbf{Benchmark} &  \textbf{Slab-Virtual} & \textbf{Direct-Mitigation} \\
        \midrule
        Sys-RAM (MB/s)          & 2.26\% & -0.43\%  \\
        Sys-CPU (Events/s)       & -0.95\% & -4.08\%   \\
        FFmpeg (s)         & -0.95\% & 0.11\% \\
        OpenSSL (Verify/s)          & 3.43\% & -2.72\%   \\
        OpenSSL (Sign/s)          & 2.78\% & 1.18\%  \\
        PHPBench (Score)    & 3.33\% & 0.40\%  \\
        PyBench (ms)    & 11.70\% & -1.29\%   \\
        GIMP (s)      & 10.19\% & 5.57\% \\
        PostMark (TPS)      & 7.67\% & 1.00\% \\
        Apache (Req/s)  & 3.80\% & 1.43\% \\
        Memcached(Ops/s) & 10.64\% & 1.93\% \\
        Redis(Req/s) & 3.68\% & -0.82\% \\
        Nginx(Req/s) & -3.14\% & -1.62\% \\
        \midrule
        Geo-Mean & 4.09\% & 0.02\% \\
        \bottomrule
        
    \end{tabular}
    \caption{The Overhead on Phoronix Benchmarks.}
    \vspace{-2ex}
    \label{tab:phoronixbench}
\end{table}

%% file: sections/10-related_work.tex
\section{Related Works}


This paper primarily focuses on conducting a systematic study for an exploitation technique for the Linux kernel exploitation. In this section, we provide an overview of these related works and highlight the distinctions between their approaches and the scope of our paper.


\paragraph{Exploitability Assessment \& Improvement.} FUZE~\cite{fuze}, aims to strengthen the evaluation of kernel UAF vulnerability exploitability and guide the manipulation of vulnerable object, by combining kernel fuzzing along with symbolic execution. KOOBE~\cite{koobe} emphasizes the automated exploit generation (AEG) on Linux Kernel Heap Out-Of-Bound vulnerabilities, including a novel capability-guided technique, to achieve more comprehensive analysis for the capability of real-world kernel OOB vulnerabilities. These two works are state-of-the-art in automatic generation of traditional kernel heap exploits, however, they are different from our work, which is a comprehensive study on a lesser known technique itself. ExpRace~\cite{exprace} provides the insight of augmenting time-windows to improve the success rate on race condition vulnerability exploitation, by performing inter-process interrupts. PSPRAY~\cite{pspray} creatively proceeds to utilize a timing side-channel attack based technique, increasing the success probability of exploitation. Xu et al.~\cite{wenxu} conducts a systematic study on how to exploit use-after-free vulnerabilities in Linux kernel based on the difficulties that mainly come from the uncertainty of the kernel memory layout. Target on kernel vulnerable object, ELOISE~\cite{eloise} clarifies the exploitability of object with adjustable size, and how it facilitates the kernel vulnerability exploiation. SLAKE~\cite{slakes} benefits the community through constructing a kernel object database for heap-based vulnerabilities, and demonstrates its capability of memory layout manipulation. KEPLER~\cite{kepler} points out the communication between kernel space and user space can be used for kernel stack ROP, by stack canary leakage and payload injection. KHEAPS~\cite{kheaps} retrieves several traditional heap-based exploitation techniques, and builds combination-exploits from them to comprehensively measure the difference of exploitation stability. We consider KHEAPS~\cite{kheaps} as the most related state-of-the-art work currently, and had a systematic comparison with it in Section~\ref{sec:evaluation}.

\paragraph{Exploitation Approaches \& Techniques.} ret2usr~\cite{ret2usr} mentions the importance of isolation between the user space and kernel space, to circumvent the inadequate mitigation, kGuard is introduced as a insight to counter it. ret2dir~\cite{ret2dir}, making a constructive move, elaborates to us how implicit memory sharing can be maliciously leveraged to compromise the isolation technique.  DirtyCred~\cite{dirtycred} proposes a new and general exploitation method. They concentrate on launching attacks by Linux Kernel's privileged object which can be artfully used to replace a vulnerable object and fulfill the privilege escalation. Reshetova et al.~\cite{ebpfjitspray} retrieves the kernel eBPF JIT Spray exploitation. While they develop two different JIT Spray attack approaches by eBPF subsystem, and prove that the existing measures implemented in the upstream kernel are not enough to stop JIT Spray attack. He et al.~\cite{crosscontainerattack} involves malicious eBPF program in kernel space to break container boundary, through hijacking the user space process which has been granted to high privilege. Kirzner et al.~\cite{ebpfspec} dwells on the speculative type confusion vulnerabilities in Linux Kernel eBPF subsystem, and conducts a study on existing general mitigations, including system-level, compiler-level, and hardware level. 


In essence, our work stands apart from the aforementioned studies. While those studies focus on kernel heap-based exploitation or side-channel attacks, our work advances further by concentrating on page-level control. To the best of our knowledge, although page-level spraying technique is not a novel exploitation method, our work is the first to systematically investigate and examine it. Through real-world case studies, we aim to showcase the potential and effectiveness of this approach.

%% file: sections/11-conclusion.tex
\section{Conclusion}


In conclusion, our systematic study reveals that Page Spray as a page-level exploitation method complements existing techniques in the field. Page Spray offers a viable alternative with comparable even superior exploitability and stability in real-world scenarios. Notably, in terms of compatibility, Page Spray exhibits greater versatility, serving as an easily adaptable method during exploitation. The general model of Page Spray, \sys, demonstrates compatibility with multiple variants, underscoring its flexibility. Our investigation into the root causes of Page Spray has revealed its close association with certain mechanisms in the Linux Kernel's design. Addressing these root causes and mitigating Page Spray should be a priority for future developments. This may involve introducing enhanced memory isolation designs at the page level to enhance security measures. In essence, Page Spray emerges as a promising and powerful addition to Linux kernel exploitation techniques, offering a unique set of advantages. However, its potential risks and implications should not be underestimated, and proactive measures are warranted to safeguard against Page Spray attacks in kernel security.

%% file: sections/ack.tex
\section*{Acknowledgement}
We thank our reviewers and shepherd for their valuable feedback and comments. This work is honored to be sponsored by, and related to Google PhD Fellowship, Northwestern University PhD Fellowship, also the Advanced Research Projects Agency for Health (ARPA-H) under Contract No. SP4701-23-C-0074, the Defense Advanced Research Projects Agency (DARPA) and Naval Information Warfare Center Pacific (NIWC Pacific) under Contract No. N66001-22-C-4026 and No. N66001-20-C-4020., and Department of Navy award N00014-23-1-2563 issued by the Office of Naval Research. Any opinions, findings, and conclusions or recommendations expressed in this work are those of the author(s) and do not necessarily reflect the views of the institutions above.


%% file: sections/12-appendix.tex
\section{Appendix}
\subsection{Detail of Experiment Setup}
\label{sec:appendix1}
\paragraph{Environment used for evaluation. }
We run the experiments on a PC equipped with 12th Gen Intel Core i7-12700 @ 4.8GHz (20 cores in total) and 32GB memory, running Ubuntu 22.04 LTS. For each vulnerability, we re introduce it into v4.15 Linux kernel, compile the corresponding kernel and disk image and run the system in QEMU virtual machine (VM). The VM is configured with 2 CPUs and 2GB RAM.

\paragraph{VM vs. bare-metal machine. }
Admitted that the success rate of each exploit we obtain from VMs may be different from that on bare-metal machines, we argue that the relative difference of stability between different techniques and the change of success rate (e.g., improvement or degradation) is consistent. Therefore, by observing the results from VMs, we can safely draw conclusions.

\paragraph{Busy environment simulation. }

For the busy setting created by \code{stress-ng}, it enables a high degree of control over the type and quantity of stressors to run. In our specific experiment, we opted to engage two stressors for each of the following stressor types, including CPU, sock, shm, and timerfd. The CPU stressors consistently maintain CPU usage close to 100\%. Concurrently, the other stressors target kmalloc cache usage, affecting distinct kmalloc cache sizes. For instance, sock stressors allocate a considerable number of skbuff structs, thereby stressing kmalloc-2048. Similarly, the shm stressors allocate shminfo structs, pressuring kmalloc-64. Timerfd stressors create timerfd structs, focusing on kmalloc-256, while additional kmalloc caches also undergo stress from other structures generated by these stressors.

\subsection{Design \& Implementation}

\label{sec:appendix2}
\paragraph{Preparation. } We implemented our analyzer based on LLVM. The analysis tool  takes kernel bitcode as input. To avoid the bitcode from being optimized and lose critical dataflow information, we used a customized clang to generate bitcode before any code optimization is invoked. The first thing we perform in early stage is to scan every modules and build the global context, in this phase, we collect related global level object, structs, and functions information. Specifically, when we found certain member in a data structure, we go through each field of it, and check if it is a related subsystem's mmap function, we record and insert into a mmap-function-list. We also build the invocation information in this phase, which maintains a global level call graph. 
\paragraph{Cross Analysis.} As we mentioned in Copy-Write call model in Section~\ref{call-write-call-model}. A copy-write callsite is constructed with an allocation and a copy. Due to this, we perform a cross tracking from the allocation root interfaces and copy root interfaces. This process works based on the previous collected invocation information. Then, two invocation chains are built for the Copy-Write callsite. After we idnetify the chains, we analyze whether there are any intersections between these two call chains. Assume no intersections are found, we exclude the case and mark as invalid for copy-write type page spray. If an intersection is found we mark as potential copy-write callsite. For those potential callsites, we apply the corresponding dataflow analysis to them. The intention of dataflow analysis at this time is to check if they two chains can be traced back to the same control structure. In two chains' intersection function, we apply a forward analysis to check the control struct of the presence spot in intersection function(the instruction in the intersection function for allocation) is able to reach the corresponding position in the instruction function for data copy.

\paragraph{Remapping Analysis.} When it comes to analyze remapping callsites, it is worth to emphasize that the biggest difference compared to copy-write callsites is that the allocation chains and remapping chains probably don't have an intersection, which require more detailed dataflow analysis. First, we also build invocation chain for remapping(mmap) from root interfaces. In the following step, we search from the lowest level page allocation function with a BFS queue, to find if a potential page allocation can be invoked in the corresponding remapping's subsystem. Once we successfully find an upper root interface in subsystem, we mark it as potential allocation root interface for previous remapping chain. After that, we perform backward dataflow analysis to remapping point, the goal of it is to find a potential control structure for remapping behaviors in related subsystem. Correspondingly, to find out the control structure for allocation, we execute forward dataflow analysis. By integrating the results from the previous dataflow analysis. We are able to have a allocation control structure and remapping control structure. At last, we apply a nest-analysis for two structure to figure out whether one of the control structures is a childfield of another control structure. We confirm as valid remapping point when they have the same control structure in the same subsystem, or the two control structures have a parent-child relationship in the same subsystem.

\paragraph{Technical Discussion.} By using analyzer, we identify 21 callsites for page spray based on our invocation model. We compare these callsites with those randomly sampled and manually confirmed, and find our manually audited callsites are a subset of those pinpointed by analyzer. We understand this discovery cannot directly conclude zero false negatives because the kernel’s scale limits our ability to manually audit all potential callsites. However, it implies the false negatives of analyzer are minimal under the potential callsites of our invocation model. To provide more systematic understanding, we need to calrify that if there are some other callsites out of our current callsite model, it could also lead to extra false negatives. On the other hand, for false positive, we perform dynamic test to verify and trigger them by Syzkaller in Section~\ref{sec:syzkaller-test} , and illustrate some of them have related limitation in Table~\ref{tab:callsites}.

\subsection{A basic step-by-step Page Spray exploit}
\label{sec:appendix3}
In this Appendix Section, we will demonstrate a detailed step-by-step exploit for one of the CVEs that we used in our evaluation as an example for a generic Page Spray exploit against a UAF vulnerability. The PoCs for other CVEs follow the exact principles that we explain in this section, albeit for a vulnerability in a different subsystem. The CVE we choose to demonstrate is CVE-2018-6555, which is a UAF vulnerability in the IrDA subsystem of the Linux kernel. We will describe in details the steps that are needed to achieve the kernel heap layout similar to the one introduced in Section \S\ref{sec:expmodel}. However, the internals of the IrDA subsystem are out of the scope of this paper and will not be explained in details.

\paragraph{The vulnerability.} The \code{irda_setsockopt} function in \code{net/irda/af_irda.c} and later in \code{drivers/staging/irda/net/af_irda.c} in the Linux kernel before 4.17 allows local users to cause a UAF on the \code{ias_object} object. In the public PoC that we collected, this UAF is triggered by first creating 3 IrDA sockets using \code{socket(AF_IRDA, SOCK_STREAM, 0)} and bind them using the function \code{irda_bind}. Subsequently, calling the function \code{irda_set_ias(fd, "\\x00")} on 2 of the 3 sockets will corrupt the \code{ias_object} queue and later trigger a UAF when closing the 3 sockets. Finally, creating and binding a 4th socket will overwrite the target pointer, and calling \code{setsockopt} with the right arguments will execute the function pointer that has been overwritten.

\paragraph{Initial setup.} Before setting up the kernel heap for Page Spray and for triggering the vulnerability, we utilized some commonly used technique to make the exploit more stable. Firstly, we pin the process to a specific CPU, so that all our objects will be allocated from a single CPU. This is a requirement for a slab to be discarded after all objects inside it are freed. Secondly, we defragment the heap by allocating many \textit{padding objects}, in this case, we chose the \textit{padding objects} to be the \code{msg_msg} objects. This step is not always necessary, but it helps against very busy environments. Thirdly, we prepare beforehand the \textit{page spraying callsite} that we will use to reclaim the vulnerable page. This callsite is chosen amongst the callsites that we found in Section \S\ref{sec:analysis-result}. In most of our PoCs, we choose the \code{pipe_write} callsite for generality and ease of implementation, therefore, in this step, we prepare multiple \code{pipe} objects that we can later write to, in order to spray page allocations.

\paragraph{Page Spray setup.} Since this is a UAF vulnerability, our \textit{vulnerable object} and \textit{victim object} can be the same object, which in this case is the \code{ias_object}. As mentioned above, we choose our \textit{padding objects} to be \code{msg_msg}. Since \code{ias_object} resides in kmalloc-96, we can spray \code{msg_msg} objects of size 48 (to account for the extra 48-byte header of \code{msg_msg} objects). With the Page Spray objects in mind, these are the steps to create a kernel heap layout similar to the one we introduced in Section \S\ref{sec:expmodel} for CVE-2018-6555:

\begin{enumerate}[leftmargin=*, itemsep=1pt, topsep=1pt, partopsep=1pt, parsep=1pt]
\item Spraying our first set of \code{msg_msg} \textit{padding objects}.
\item Creating and binding 3 IrDA sockets using \code{socket(AF_IRDA, SOCK_STREAM, 0)} and \code{irda_bind}.
\item Reinserting the middle \code{ias_object} and corrupting the queue by calling the function \code{irda_set_ias(fd, "\\x00")} on 2 of the 3 sockets.
\item Spraying our second set of \code{msg_msg} \textit{padding objects}. At this point, we have achieved a vulnerable \code{ias_object} in the middle of 2 sets of \code{msg_msg} padding objects.
\item Closing all 3 sockets to trigger a free on the vulnerable \code{ias_object}.
\item Freeing all 2 sets of \code{msg_msg} padding objects. At this point, the slab containing the vulnerable \code{ias_object} will be discarded.
\end{enumerate}

\paragraph{Page Reclaim.} To reclaim the page containing the vulnerable/victim object, we simply write to the pipes that we have prepared in the initial setup step. The content of the data that we use to write will be an array of 8-byte values that are all equal to \code{0xffffffffdeadbeef}. This way, we can recognize a successful Page Spray exploit be parsing the crash report, i.e. a successful exploit will crash at RIP equals to \code{0xffffffffdeadbeef}.

\paragraph{Trigger UAF.} After corrupting the victim \code{ias_object}, we can execute one of if its function pointer by creating and binding a 4th IrDA socket, and calling \code{setsockopt}. The corrupted pointer will be executed and the kernel will crash at RIP equals to \code{0xffffffffdeadbeef}, resulting in a successful Page Spray exploit and hijack the kernel's execution flow.

%% file: usenix.bbl
\begin{thebibliography}{10}

\bibitem{SLAB}
{Chapter 8 Slab Allocator}.
\newblock \url{https://www.kernel.org/doc/gorman/html/understand/understand011.html}.

\bibitem{koobe}
Weiteng Chen, Xiaochen Zou, Guoren Li, and Zhiyun Qian.
\newblock {KOOBE}: Towards facilitating exploit generation of kernel {Out-Of-Bounds} write vulnerabilities.
\newblock In {\em Proceedings of the 29th {USENIX} Security Symposium}, 2020.

\bibitem{elastic}
Yueqi Chen, Zhenpeng Lin, and Xinyu Xing.
\newblock A systematic study of elastic objects in kernel exploitation.
\newblock In {\em Proceedings of the 2020 ACM SIGSAC Conference on Computer and Communications Security}, 2020.

\bibitem{eloise}
Yueqi Chen, Zhenpeng Lin, and Xinyu Xing.
\newblock A systematic study of elastic objects in kernel exploitation.
\newblock In {\em Proceedings of the 2020 ACM SIGSAC Conference on Computer and Communications Security}, 2020.

\bibitem{slakes}
Yueqi Chen and Xinyu Xing.
\newblock Slake: Facilitating slab manipulation for exploiting vulnerabilities in the linux kernel.
\newblock In {\em Proceedings of the 2019 ACM SIGSAC Conference on Computer and Communications Security}, 2019.

\bibitem{partitionalloc}
chromium.
\newblock {PartitionAlloc Design}.
\newblock \url{https://chromium.googlesource.com/chromium/src/+/master/base/allocator/partition\_allocator/PartitionAlloc.md}.

\bibitem{SLUB}
Corbet.
\newblock {The SLUB allocator}.
\newblock \url{https://lwn.net/Articles/229984/}, 2007.

\bibitem{CVE-2022-2585}
{CVE-2022-2585}.
\newblock \url{https://ssd-disclosure.com/ssd-advisory-linux-clock\_thread\_cputime\_id-lpe/}, 2022.

\bibitem{CVE-2022-2588}
{CVE-2022-2588}.
\newblock \url{https://github.com/Markakd/CVE-2022-2588}, 2022.

\bibitem{OOB-read}
{CWE-125: Out-of-bounds Read}.
\newblock \url{https://cwe.mitre.org/data/definitions/125.html}.

\bibitem{OOB-write}
{CWE-125: Out-of-bounds Write}.
\newblock \url{https://cwe.mitre.org/data/definitions/787.html}.

\bibitem{DF}
{CWE-415: Double Free}.
\newblock \url{https://cwe.mitre.org/data/definitions/415.html}.

\bibitem{UAF}
{CWE-416: Use After Free}.
\newblock \url{https://cwe.mitre.org/data/definitions/416.html}.

\bibitem{kaslr}
Jake Edge.
\newblock {Kernel address space layout randomization. (2013)}.
\newblock \url{https://lwn.net/Articles/569635/}, 2013.

\bibitem{cfi}
Jake Edge.
\newblock {Control-flow integrity for the kernel.}
\newblock \url{https://lwn.net/ Articles/569635/}, 2020.

\bibitem{freelistrandom}
Thomas Garnier.
\newblock {mm: SLAB freelist randomization}.
\newblock \url{https://lwn.net/Articles/685047/}, 2016.

\bibitem{kcfi}
Google.
\newblock {Kernel Control Flow Integrity}.
\newblock \url{https://source.android.com/docs/security/test/kcfi}.

\bibitem{syzbot-assets}
Google.
\newblock {syzbot-assets}.
\newblock \url{https://storage.googleapis.com/syzbot-assets/067efcb9fd9a/ci-qemu-upstream-e8f897f4.html#drivers%2fnet%2ftap.c}.

\bibitem{syzkaller}
Google.
\newblock syzkaller \- kernel fuzzer, 2022.

\bibitem{crosscontainerattack}
Yi~He, Roland Guo, Yunlong Xing, Xijia Che, Kun Sun, Zhuotao Liu, Ke~Xu, and Qi~Li.
\newblock Cross container attacks: The bewildered {eBPF} on clouds.
\newblock In {\em Proceedings of the 32rd {USENIX} Security Symposium}, 2023.

\bibitem{smep}
Mateusz Jurczyk and Gynvael Coldwind.
\newblock {SMEP: What is it, and how to beat it on Windows}.
\newblock \url{https://j00ru.vexillium.org/2011/06/smep-what-is-it-and-how-to-beat-it-on-windows/}, 2011.

\bibitem{kCTF}
{kCTF is a Kubernetes-based infrastructure for CTF competitions}.
\newblock \url{https://google.github.io/kctf/}.

\bibitem{ret2dir}
Vasileios~P. Kemerlis, Michalis Polychronakis, and Angelos~D. Keromytis.
\newblock Ret2dir: Rethinking kernel isolation.
\newblock In {\em Proceedings of the 23rd {USENIX} Security Symposium}, 2014.

\bibitem{ret2usr}
Vasileios~P. Kemerlis, Georgios Portokalidis, and Angelos~D. Keromytis.
\newblock {kGuard}: Lightweight kernel protection against {Return-to-User} attacks.
\newblock In {\em Proceedings of the 21st {USENIX} Security Symposium}, 2012.

\bibitem{skb}
The kernel~development community.
\newblock {struct sk\_buff}.
\newblock \url{https://docs.kernel.org/networking/skbuff.html}.

\bibitem{ebpfspec}
Ofek Kirzner and Adam Morrison.
\newblock An analysis of speculative type confusion vulnerabilities in the wild.
\newblock In {\em Proceedings of the 30th {USENIX} Security Symposium}, 2021.

\bibitem{pspray}
Yoochan Lee, Jinhan Kwak, Junesoo Kang, Yuseok Jeon, and Byoungyoung Lee.
\newblock Pspray: Timing {Side-Channel} based linux kernel heap exploitation technique.
\newblock In {\em Proceedings of the 32rd {USENIX} Security Symposium}, 2023.

\bibitem{exprace}
Yoochan Lee, Changwoo Min, and Byoungyoung Lee.
\newblock {ExpRace}: Exploiting kernel races through raising interrupts.
\newblock In {\em Proceedings of the 30th {USENIX} Security Symposium}, 2021.

\bibitem{autoslab}
Zhenpeng Lin.
\newblock {AUTOSLAB}.
\newblock \url{https://grsecurity.net/how\_autoslab\_changes\_the\_memory\_unsafety\_game}.

\bibitem{dirtycred}
Zhenpeng Lin, Yuhang Wu, and Xinyu Xing.
\newblock Dirtycred: Escalating privilege in linux kernel.
\newblock In {\em Proceedings of the 2022 ACM SIGSAC Conference on Computer and Communications Security}, 2022.

\bibitem{badiouring}
Zhenpeng Lin, Xinyu Xing, Zhaofeng Chen, and Kang Li.
\newblock Bad io\_uring: A new era of rooting for android.
\newblock In {\em Blackhat USA}, 2023.

\bibitem{linux-selftest}
Linux.
\newblock {Linux Kernel Selftests}.
\newblock \url{https://www.kernel.org/doc/html/v6.1/dev-tools/kselftest.html}.

\bibitem{cross-cache1}
{Linux Kernel PWN | 06 DirtyCred}.
\newblock \url{https://blog.wohin.me/posts/linux-kernel-pwn-06/}.

\bibitem{SLOB}
Matt Mackall.
\newblock {slob: introduce the SLOB allocator}.
\newblock \url{https://lwn.net/Articles/157944/}, 2005.

\bibitem{kpti}
{Page Table Isolation (PTI)}.
\newblock \url{https://www.kernel.org/doc/html/next/x86/pti.html}.

\bibitem{phoronixbenchmark}
{Phoronix Test Suite}.
\newblock \url{https://www.phoronix-test-suite.com/}.

\bibitem{freelistharden}
Alexander Popov.
\newblock {Add SLUB free list pointer obfuscation}.
\newblock \url{https://samsung.github.io/kspp-study/heap-ovfl.html}, 2017.

\bibitem{ebpfjitspray}
Elena Reshetova, Filippo Bonazzi, and N.Asokan.
\newblock {Randomization can’t stop BPF JIT spray}.
\newblock In {\em Blackhat EU}, 2016.

\bibitem{cross-cache2}
{Reviving Exploits Against Cred Structs}.
\newblock \url{https://www.willsroot.io/2022/08/reviving-exploits-against-cred-struct.html}.

\bibitem{stressng}
{stress-ng - a tool to load and stress a computer system}.
\newblock \url{https://manpages.ubuntu.com/manpages/jammy/man1/stress-ng.1.html}.

\bibitem{smap}
{Supervisor mode access prevention}.
\newblock \url{https://lwn.net/Articles/517475/}, 2012.

\bibitem{RCU}
{What is RCU? -- "Read, Copy, Update"}.
\newblock \url{https://www.kernel.org/doc/html/next/RCU/whatisRCU.html}.

\bibitem{kepler}
Wei Wu, Yueqi Chen, Xinyu Xing, and Wei Zou.
\newblock {KEPLER}: Facilitating control-flow hijacking primitive evaluation for linux kernel vulnerabilities.
\newblock In {\em Proceedings of the 28th {USENIX} Security Symposium}, 2019.

\bibitem{fuze}
Wei Wu, Yueqi Chen, Jun Xu, Xinyu Xing, Xiaorui Gong, and Wei Zou.
\newblock Fuze: Towards facilitating exploit generation for kernel use-after-free vulnerabilities.
\newblock In {\em Proceedings of the 27th {USENIX} Security Symposium}, 2018.

\bibitem{wenxu}
Wen Xu, Juanru Li, Junliang Shu, Wenbo Yang, Tianyi Xie, Yuanyuan Zhang, and Dawu Gu.
\newblock From collision to exploitation: Unleashing use-after-free vulnerabilities in linux kernel.
\newblock In {\em Proceedings of the 2015 ACM SIGSAC Conference on Computer and Communications Security}, 2015.

\bibitem{kheaps}
Kyle Zeng, Yueqi Chen, Haehyun Cho, Xinyu Xing, Adam Doup{\'e}, Yan Shoshitaishvili, and Tiffany Bao.
\newblock Playing for {K(H)eaps}: Understanding and improving linux kernel exploit reliability.
\newblock In {\em Proceedings of the 31st {USENIX} Security Symposium}, 2022.

\bibitem{mte}
Google~Project Zero.
\newblock {First handset with MTE on the market}.
\newblock \url{https://googleprojectzero.blogspot.com/2023/11/first-handset-with-mte-on-market.html}.

\end{thebibliography}
